\documentclass[nodate,showpacs,amsmath,amssymb,aps,prl]{revtex4-1}
\usepackage{color}

\def\qt{\ln q^{\frac{1}{3}}}
\def\MC{\mathrm{M}}
\def\S{\mathcal{S}}
\def\3G{^{(3)}\mathcal{G}}

\begin{document}

\title{General Relativity without paradigm of space-time covariance, and resolution of the problem of time}

\author{Chopin Soo}\email{cpsoo@mail.ncku.edu.tw}
\affiliation{Department of Physics, National Cheng Kung University, Taiwan}
\author{Hoi-Lai Yu}\email{hlyu@phys.sinica.edu.tw}
\affiliation{Institute of Physics, Academia Sinica, Taiwan}
\begin{abstract}
The framework of a theory of gravity from the quantum to the classical regime is presented. The paradigm shift from full spacetime covariance to spatial diffeomorphism invariance, together with clean decomposition of the canonical structure, yield transparent physical dynamics and a resolution of the problem of time. The deep divide between quantum mechanics and conventional canonical formulations of quantum gravity is overcome with a Schr\"{o}dinger equation for quantum geometrodynamics that describes evolution in intrinsic time. Unitary time development with gauge-invariant temporal ordering is also viable.  All Kuchar observables become physical; and classical spacetime, with direct correlation between its proper times and intrinsic time intervals, emerges from constructive interference. The framework not only yields a physical Hamiltonian for Einstein's theory, but also prompts natural extensions and improvements towards a well behaved quantum theory of gravity. It is a consistent canonical scheme to discuss Horava-Lifshitz theories with intrinsic time evolution, and of the many possible alternatives that respect 3-covariance (rather than the more restrictive 4-covariance of Einstein's theory), Horava's ``detailed balance" form of the Hamiltonian constraint is essentially pinned down by this framework.  Issues in quantum gravity that depend on radiative corrections and the rigorous definition and regularization of the Hamiltonian operator are not addressed in this work.
\end{abstract}
\pacs{04.60.-m, 04.60.Ds\\ 
Published: {\bf Prog. Theor. Exp. Phys. (2014) 013E01} (doi:10.1093/ptep/ptt109)}

\maketitle

\section{1. Introduction and Overview}\label{sec1}

 Covariance of space and time has been crucial to Einstein's theory of General Relativity (GR); but the assumption of full 4-dimensional (4D)  diffeomorphism symmetry also entails a number of technical and conceptual difficulties. On the technical front, Einstein's theory fails to be renormalizable as a perturbative quantum field theory; and GR remains incomplete as a quantum theory, despite many recent advances. At  the conceptual level, the Hamiltonian constraint, which is believed to dictate the dynamics has a dual role as a generator of symmetry, with an arbitrary lapse function as the Lagrange multiplier. Full 4D invariance with a local Hamiltonian constraint and its consequent baggage of arbitrary lapse and gauged histories is hard to reconcile with the physical reality of time.

Canonical quantum gravity has the guise of a formalism ``frozen in time". Quantum states do not evolve in coordinate ``time", and $dx^0$  is merely one component of spacetime displacement, with no invariant physical meaning in a theory with full spacetime covariance. Yet a notion of time is needed to make sense of quantum mechanical interpretations, both for the discussion of dynamics and evolution, and for the interpretation of probabilities that are normalized at particular instants of time. Even if a suitable degree of freedom (d.o.f.) is chosen as the intrinsic time (as many have advocated), a successful quantum theory will still need to reconcile a Klein-Gordon type Wheeler-DeWitt (WDW) equation\cite{DeWitt, Wheeler} quadratic in momenta with the positivity of ``probabilities". A resolution of the ``problem of time"(see, e.g. Refs.\cite{Isham, Kuchar2, Anderson}) cannot be deemed complete if it fails to account for the intuitive physical reality of time and does not provide a satisfactory correlation between intrinsic time development in quantum dynamics and the passage of time in classical spacetimes.

At the deeper level of symmetries, the constraints of GR satisfy the Dirac algebra\cite{Dirac}. However, closer inspection reveals it is not the algebra of the generators of 4D diffeomorphisms.
 Off-shell, full 4D Lie derivatives involve time derivatives; these cannot be generated by constraints that depend only on the spatial metric  $q_{ij}$ and its conjugate momentum ${\tilde\pi}^{ij}$. In Einstein's theory, the equations of motion (EOM), among other things, relate $\dot {q}_{ij}$ to ${\tilde \pi}^{ij}$ and the constraints then generate 4D diffeomorphisms on-shell (see, e.g. Ref.\cite{TThiemann}). This is a clue that, without the help of EOM, a {\it quantum} theory of GR cannot, and need not, enforce full 4D spacetime covariance {\it off-shell}.

A key obstacle to the viability of GR as a perturbative quantum field theory lies in the conflict between unitarity and spacetime general covariance: renormalizability can be attained with higher-order curvature terms, but spacetime covariance requires time as well as spatial derivatives of the same (higher) order, thus compromising unitarity. Horava relinquished full 4D symmetry and achieved power-counting renormalizable modifications of GR with higher-order spatial curvature terms\cite{Horava}.  In loop quantum gravity, the non-perturbative master constraint program\cite{Thiemann} seeks representations not of the Dirac algebra, but of the master constraint algebra which has the advantages of having structure {\it constants} (rather than functions), and of decoupling the equivalent quantum Hamiltonian constraint from spatial diffeomorphism generators $H_i$. Reference \cite{Soo and Yu} consistently realized  Horava gravity theories as canonical theories with first-class master constraint algebra. This not only removes the canonical inconsistencies of projectable Horava theory, but also captures the essence of the theory in retaining spatial diffeomorphisms as the {\it only} local gauge symmetries. With full covariance, observables, $O$, of Einstein's theory are required to commute both with $H_i$ {\it and} $H$, leading to the demand (above and beyond the usual canonical rules) of $\{O, \{O,\mathrm{M}\}\}|_{\mathrm{M}=0} = 0$\cite{Thiemann}. In contradistinction, in this work a theory of gravity appositely formulated with a master constraint marks a real paradigm shift in the symmetry, from full 4D general coordinate invariance to invariance only with respect to spatial diffeomorphisms. Enforcing $H(x)=0$ through the master constraint effectively removes it from one of its dual roles since $\mathrm{M}$ generates no extra symmetry, and determines only dynamics.

The call to abandon 4-covariance is not new. In the simplification of the Hamiltonian analysis of GR, the fact that only the spatial metric is physical led Dirac to conclude that ``four-dimensional symmetry is not a fundamental property of the physical world"\cite{Dirac2}. In his seminal article, Wheeler emphasized that spacetime is a concept of ``limited applicability", and it is 3-geometry, rather than 4-geometry, that is fundamental in quantum geometrodynamics\cite{Wheeler}. That semiclassical spacetime is emergent begs the question what, if anything at all, takes the place of  ``time" in quantum gravity? Wheeler went as far as to claim we have to forgo time-ordering, and to declare that ``there is no spacetime, there is no time, there is no before, there is no after"\cite{Wheeler}.  But without ``time -ordering", how can ``causality",  which is requisite in any ``sensible physical theory", be ensured in quantum gravity?  Quantum geometrodynamics, it will be revealed in this work, is dictated by Schr\"{o}dinger evolution with respect to intrinsic time; and quantum gravity can be formulated as a causal quantum field theory with diffeomorphism-invariant temporal ordering.

A theory of geometrodynamics, with $(q_{ij}, {\tilde \pi}^{ij})$ as fundamental variables, is bequeathed with a number of remarkable features: positivity of the spatial metric guarantees space-like separation between any two points on the initial value hypersurface (and allows, to quote Ref.\cite{Wheeler}, ```a notion of `simultaneity' and a common moment of a rudimentary `time'"). In the Arnowitt-Deser-Misner (ADM)\cite{ADM} description of the spacetime metric, these are labeled as constant-$x^0$ hypersurfaces. Quantum states, however, do not depend on $x^0$; the link between intrinsic and coordinate times which will be revealed is more subtle.
Decomposition of $q_{ij}= q^{\frac{1}{3}}{\bar q}_{ij}$ into determinant and unimodular factors results in clean separation of canonical pairs.  The generic ultra-local DeWitt supermetric\cite{DeWitt} with deformation parameter $\lambda$ has the signature $({\rm sgn}[\frac{1}{3}-\lambda], +,+, + ,+ ,+)$, and the single negative eigenvalue for $\lambda > \frac{1}{3}$ corresponds to the $\delta\ln q$ mode.  It is the apposite choice (as subsequent discussions will show)  for the intrinsic time interval in quantum gravity.  In a mini-superspace context,  Misner also advocated $\ln q$ as a time variable and obtained the corresponding Hamiltonian\cite{Misner}. Although $q$ is a scalar density (a criticism that has been used to disqualify it from the role of time variable), $\delta\ln q= \frac{\delta q}{q}$ is a spatial diffeomorphism scalar. Even in Galilean-Newtonian physics, it is {\it time interval}, and not absolute time, which is physical.

With the preceding concerns and observations in mind, the framework for a theory of gravity, from the quantum to the classical regime, will be presented wherein several factors collude to result in a minimal and compelling formulation. In Sect. 2, it will be demonstrated that  the WDW constraint permits factorization; and the classical content of GR may be captured by $({\beta\pi}+{\bar H})=0$ with $\beta^2 =\frac{1}{6}$. That  $\pi$, the trace of the momentum, is conjugate to $\qt$ leads to a quantum theory described by a Schr\"{o}dinger equation {\it first order in intrinsic time}, with consequent positive semidefinite probability density $\Psi^\dagger\Psi$. The resultant semiclassical Hamilton-Jacobi (HJ) equation is also of first order, with the implication of completeness\cite{Courant, Landau} (its integral solution provides a complete set of gauge-invariant integration constants of motion). Classical spacetime emerges from constructive interference of quantum wave functions\cite{Gerlach}. The physical content of GR is regained from a theory with ${\bar H}/{\beta}$ generating intrinsic $\delta\qt$ time translations, subject to only spatial diffeomorphism invariance. Furthermore, the emergent classical spacetimes have ADM metrics with proper times (for vanishing shifts) directly correlated to intrinsic time intervals by the explicit relation $d\tau^2 = [\frac{\delta\ln q}{(12\beta\kappa {\bar H}/{\sqrt q})}]^2$. These results are derived in Sects. 2.2 and 2.3; and the master constraint formulation is introduced in Sect. 2.1. There is the additional promise of a well behaved unitary quantum theory of gravity if the Hamiltonian generating intrinsic time development can be made real, bounded from below, and also renormalizable by including suitable higher-order spatial curvature terms (with GR recovered in the limit of low curvatures). These modifications are discussed in Sect. 2.4. The emergence of a {\it global} intrinsic time parameter, $h$, from the fundamental equation governing quantum geometrodynamics in superspace, $i\hbar\frac{\delta \Psi}{\delta h} =\left[\int\frac{{\bar H}(x)}{\beta}d^3x\right]\Psi$, is addressed in Sect. 2.5, together with {\it gauge-invariant temporal ordering} in the Heisenberg picture with unitary $h$-ordered evolution operator. Section 3 contains further discussions on the physical degrees of freedom and the relation of this framework to other works.

It must be pointed out that $\bar H$ involves a square root and its rigorous definition remains a formidable challenge. Issues in quantum gravity that require the rigorous definition and regularization of the Hamiltonian operator are thus not yet addressed in the current work.

\section{2. Theory of gravity without the paradigm of full spacetime covariance}
In the initial value problem, York\cite{York1, York2} explored a conformal decomposition of the spatial metric
$
q_{ij} = \phi \bar{q}_{ij},
$
of which $\phi = q^{\frac{1}{3}}$  with unimodular $\bar{q}_{ij}$  is a special case ($q:=\det[q_{ij}]$). However, in focusing on the generic conformal factor $\phi$, the ``miracle" of  $\phi = q^{\frac{1}{3}}$ was not fully revealed. Among other advantages, it results in a clean separation of $(\qt, \pi)$ from other canonical pairs, with the symplectic potential
\begin{equation}
 \int\, {\tilde\pi}^{ij}\delta q_{ij} = \int {\bar\pi}^{ij}\delta{\bar q}_{ij} + \pi \delta \qt,
\end{equation}
wherein $\pi := q_{ij}\tilde{\pi}^{ij}$  and $ \bar{\pi}^{ij}:= q^{\frac{1}{3}}   [\tilde{\pi}^{ij}  - \frac{q^{ij}}{3}{\pi}] $. Thus the only nontrivial Poisson brackets are
$
    \{\bar{q}_{kl}(x),\bar{\pi}^{ij}(x')\}=P^{ij}_{kl}\,\delta(x,x')$, and
$        \{\ln q^{\frac{1}{3}}(x),\pi(x')\}=\delta(x,x');
$
with the trace-free projector, $P^{ij}_{kl} :=  \frac{1}{2}(\delta^i_k\delta^j_l + \delta^i_l\delta^j_k) - \frac{1}{3}\bar{q}^{ij}\bar{q}_{kl}$. This separation carries over to the quantum theory, and permits a d.o.f., separate from the others, to be identified as the carrier of temporal information.  An intrinsic clock is in fact not tied to 4D general covariance. A generic ultra-local DeWitt supermetric\cite{DeWitt} compatible with spatial diffeomorphism invariance,
$
G_{ijkl} = \frac{1}{2}(q_{ik}q_{jl} +  q_{il}q_{jk}) - \frac{\lambda}{3\lambda -1} q_{ij}q_{kl}
 $,
comes equipped with intrinsic temporal intervals $\delta\qt$  provided $\lambda > \frac{1}{3}$.
 Crucially it is also $\qt$ which can be so neatly separated from the rest.
With $\beta^2 := \frac{1}{3(3\lambda - 1)}$, the Hamiltonian constraint of a theory of geometrodynamics quadratic in momenta and with ultra-local supermetric is generically,
\begin{eqnarray}\label{GeneralH}
0=\frac{\sqrt{q}}{2\kappa} H =  G_{ijkl}\tilde{\pi}^{ij}\tilde{\pi}^{kl} + V(q_{ij})
=-(\beta \pi- \bar{H})(\beta \pi + \bar{H});\nonumber \\
\bar {H}(\bar{\pi}^{ij}, \bar {q}_{ij}, q)  = \sqrt{ \bar{G}_{ijkl}\bar{\pi}^{ij}\bar{\pi}^{kl} +  V(\bar {q}_{ij}, q) }
= \sqrt{ \frac{1}{2}[\bar{q}_{ik}\bar{q}_{jl} +\bar{q}_{il}\bar{q}_{jk}]\bar{\pi}^{ij}\bar{\pi}^{kl} + V({q}_{ij})}.
\end{eqnarray}
Einstein's GR ($\lambda = 1$ and $V({\bar q}_{ij}, q) =- \frac{q}{(2\kappa)^2}[R - 2\Lambda_{\it{eff}} $]) is a particular realization of a wider class of theories.

 As written, Eq. (\ref{GeneralH}) is a local constraint, in addition to those of spatial diffeomorphism invariance $H_i=0$.
This leads to a quandary: if the constraint algebra is first class, then $\int NH$ consistently generates multi-fingered time-translation {\it symmetry} as well as ``physically" evolving the theory with respect to ``time" (as manifested by its dual roles in Einstein's theory with full general coordinate invariance).

\subsection{2.1 Master constraint formulation}
 A master constraint formulation can equivalently enforce the local content of (\ref{GeneralH}). This permits $H$ to determine dynamical evolution rather than generate symmetry. The master constraint formulation serves as a consistent canonical method to regain the physical content of Einstein's theory, without the paradigm of 4-covariance, from the usual starting point of canonical general relativity with the Hamiltonian $\int (NH + N^iH_i)$. The analogy to simple relativistic point particle mechanics is recounted in the Appendix.
 In the master constraint formulation, the commencing action is
 \begin{eqnarray}S &=&\int [{\tilde\pi}^{ij}{{\dot q}_{ij}}- N^iH_i]d^3xdt- \int m(t)\mathrm{M}dt;\nonumber\\
  \mathrm{M}&:=&\int {({\beta\pi}+{\bar H})^2}/{\sqrt{q}}=0; \qquad H_i := -2q_{ik}\nabla_j{\tilde\pi}^{jk}.
  \end{eqnarray}
The resultant constraint algebra, $\{\mathrm{M},\mathrm{M}\}=0, \{H_i[N^i],\mathrm{M}\}=0, \{H_i[N^i],H_j[{N'}^j]\}=H_i[{\pounds}_{\vec{N}}{N' }^i]$, is first class and exhibits only spatial diffeomorphism gauge symmetry, both on- and off-shell. $\MC$  decouples from $H_i$ in the constraint algebra, thus paving the road for non-perturbative quantization\cite{Thiemann}. The result is a theory with only spatial diffeomorphism invariance; with physical dynamics dictated by $H$, but encoded in $\MC$.

Modulo $\MC=0$, the total constraints of the theory generate only spatial diffeomorphisms since $\{f(q_{ij}, {\tilde\pi}^{ij}), {m(t)}\mathrm{M} + H_k[N^k]\}|_{\mathrm{M}=0
\Leftrightarrow H=0} \approx  \{ f, H_k[N^k]\}
={\pounds}_{\vec N}f$.
Thus $m(t)$ plays no role in the dynamics. Instead, true physical evolution can only be with respect to an intrinsic time extracted from the WDW constraint.
As detailed above, $\qt$ is the preeminent choice.

The factorization allows for the possibility that $(\beta \pi + {\bar H})=0$  is sufficient to recover the classical content of GR.
This is a breakthrough: the semiclassical HJ equation is first order in intrinsic time (since $\pi$  is conjugate to $\qt$), with its consequence of completeness\cite{Courant, Landau};
and quantum gravity will now be dictated by a corresponding Schr\"{o}dinger equation of first order in intrinsic time (with consequent positive semidefinite probability density at any instant of intrinsic time). This resolves the deep divide between a quantum mechanical interpretation (in which {\it both} the notion of time and positive semidefinite probabilities are needed) and the usual Klein-Gordon type WDW equations of second order in intrinsic time (hence indefinite in ``probabilities"). In general $\beta=\pm\sqrt{\beta^2}$, but the positive value for $\beta$ is singled out to render the physical Hamiltonian ${\mathcal H}_{phys.} =\int {\bar H}/\beta$ (which shall be discussed later) positive semidefinite, which implies that $\pi$ is negative in general. In Robertson-Walker cosmological models this corresponds to an expanding universe.

 A spatial diffeomorphism-invariant quantum theory with  $\MC:= \int (\beta\pi + {\bar H})^2/\sqrt{q}=0$ will translate into
 \begin{equation}\label{WD}
[\beta\hat{\pi} + {\bar H}(\hat{\bar{\pi}}^{ij}, \hat{\bar {q}}_{ij}, \hat{q})]|\Psi\rangle=0,\,\,
{\hat{H}_i}|\Psi\rangle =0.
 \end{equation}
In the metric representation, canonical momenta are realized by $\hat{\pi} = \frac{3\hbar}{i}\frac{\delta}{\delta \ln q}$, $\hat{\bar{\pi}}^{ij} = \frac{\hbar}{i}P^{ij}_{lk}\frac{\delta}{\delta\bar{q}_{lk}}$ which operate on $\Psi[{\bar q }_{ij}, q]$, and the Schr\"{o}dinger equation and HJ equation for semiclassical states $Ce^{\frac{iS}{\hbar}}$ are respectively,
\begin{equation}\label{HJ}
i\hbar \frac{\delta}{\delta \ln q} {\Psi} =\frac{{\bar H}({\hat{\bar{\pi}}^{ij}}, q_{ij})}{3\beta} \Psi,\,
\frac{\delta S}{\delta \ln q}=-\frac{{\bar {H}}(\bar{\pi}^{ij} = P^{ij}_{kl}\frac{\delta S}{\delta{\bar  q}_{kl}}; {q}_{ij})}{3\beta};
\end{equation}
$\nabla_j\frac{\delta\Psi}{\delta q_{ij}} =0$ enforces spatial diffeomorphism symmetry.
Behold the appearance of a true Hamiltonian $\bar H/\beta$ generating evolution w.r.t. intrinsic time $\delta\qt(x)$.

 It should be emphasized that the method of master constraint is only one of the complementary approaches in our work, and the same physics can be deduced from the Schr\"{o}dinger equation above, and the Heisenberg formulation and generalized Baierlein-Sharp-Wheeler (BSW) action which shall all be discussed later. What is important is the paradigm shift to spatial diffeomorphism invariance which reveals the primacy of dynamics, with respect to intrinsic time, dictated by $\bar H$. This shift also allows consistent extensions and improvements to Einstein's theory.

\subsection{2.2 Emergence of classical spacetime}

Many years ago Gerlach\cite{Gerlach} demonstrated that classical spacetime and its EOM can be recovered from the quantum theory through HJ theory and constructive interference.
The first-order HJ equation, which bridges the quantum and classical regimes, has the complete solution $S = S(^{(3)}\mathcal{G}; \alpha)$ which depends on 3-geometry $^{(3)}\mathcal{G}$  and integration constants (denoted generically here by $\alpha$). Constructive interference with  $\S (\3G; \alpha +\delta\alpha)= \S(\3G; \alpha)$;
$\S(\3G +\delta\3G; \alpha + \delta \alpha) = \S(\3G+\delta\3G; \alpha)$ leads to
$
\frac{\delta}{\delta\alpha}\big{[}\int\,\frac{\delta\S(\3G;\alpha)}{\delta q_{ij}} \delta q_{ij} \big{]}= 0;
$
 subject to constraints $\MC =H_i=0$. With the momenta identified with ${\tilde\pi}^{ij}(\alpha) :=\frac{\delta\S(\3G;\alpha)}{\delta q_{ij}}$, and Lagrange multipliers $\delta m$ and $\delta{\texttt{N}}^i$,
the requirement of constructive interference is equivalent to
\begin{eqnarray}\label{Vary}
0=\frac{\delta}{\delta\alpha}[\int ({\tilde\pi}^{ij}\delta q_{ij} -  \delta{\texttt{N}}^i H_i )  - \delta m \MC]\nonumber\\
=\frac{\delta}{\delta\alpha}[\int \pi \delta\qt +\bar{\pi}^{ij}\delta \bar{q}_{ij}  +\frac{2q^{ij}}{3}\delta {\texttt{N}}_i\nabla_j \pi
+2 q^{-\frac{1}{3}}{\delta{\texttt{N}}_i}\nabla_j \bar{\pi}^{ij}].
\end{eqnarray}
Happily, for master constraint theories, there is no $\delta m$ contribution (since $\MC =0 \Leftrightarrow H=0$ and $\MC$ is quadratic in $H$). Imposing $\pi = -{\bar{H}}/{\beta}$, integrating by parts, and bearing in mind ${\bar H}({\bar\pi}^{ij}(\alpha), q_{ij})$, the resultant EOM is
\begin{equation}\label{qeq}
\frac{\delta \bar{q}_{ij}(x) -\pounds_{\vec{N} dt}\bar{q}_{ij}(x)}{\delta \qt (y)-\pounds_{\vec{N}dt}\qt(y) }= P^{kl}_{ij} \frac{\delta [\bar{H}(y)]}{\beta\delta \bar \pi^{kl}(x)}
=\frac{{\bar G}_{ijmn}{\bar \pi}^{mn}}{\beta{\bar H}}\delta(x,y),
\end{equation}
wherein $\pounds_{{\vec N} }$ denotes Lie derivative and $ \delta\vec{\texttt{N}} =:\vec{N}dt$.
Proceeding as in Ref.\cite{Gerlach}, the other half of Hamilton's equations,
\begin{eqnarray}\label{p}
\frac{\delta \bar{\pi}^{ij}(x) -\pounds_{\vec{N} dt}\bar{\pi}^{ij}(x)}{\delta \qt (y)-\pounds_{\vec{N}dt}\qt(y) } = -  \frac{\delta [\bar{H}(y)/\beta]}{\delta \bar q_{ij}(x)},
\end{eqnarray}
can be recovered. As predicted by (\ref{HJ}), $\bar H/\beta$ is the Hamiltonian for evolution of $({\bar q}_{ij}, {\bar\pi}^{ij})$ w.r.t. $\qt$.

Although the derivation above bears similarities to Gerlach's work, fundamental differences must be noted. In Ref.\cite{Gerlach}, $\delta{\texttt{N}} =: Ndt $ (associated with the local constraint $H=0$) will always contribute to the final EOM resulting in multi-fingered time with an arbitrary lapse function.  In contradistinction, the $\delta m$ contribution does not arise for a master constraint theory. This is part and parcel of the paradigm shift. Not only is unphysical time development with arbitrary lapse function now evaded, the ``defect" that $\MC$ does not generate dynamical evolution w.r.t. coordinate time is redeemed at a much deeper level through physical evolution w.r.t. intrinsic time. Through (\ref{qeq}) and $\pi =-{\bar H}/\beta$, the emergent ADM classical spacetime has the momentum
\begin{equation}\label{lap}
G_{ijkl}{\tilde\pi}^{kl} =\frac{\sqrt q}{4N\kappa}(\frac{dq_{ij}}{dt}-\pounds_{\vec{N}}q_{ij}),\,
{N}dt:=\frac{\delta \qt-\pounds_{\vec{N}dt}\qt}{(4\beta\kappa\bar{H}/\sqrt{q})}.
\end{equation}

In the conventional canonical formulation of Einstein's GR, the EOM  with {\it arbitrary} lapse is,
\begin{equation}\label{eq:qdot}
\frac{dq_{ij}}{dt} = \big{\{} q_{ij}, \int {N} H + N_iH^i\big{\}}
= \frac{4{N}\kappa}{\sqrt q}{G_{ijkl}}{\tilde\pi}^{kl}  + \pounds_{\vec{N}} q_{ij}.
\end{equation}
The extrinsic curvature is related to ${\tilde\pi}^{ij}$ by
$K_{ij} := \frac{1}{2N}(\frac{dq_{ij}}{dt} - \pounds_{\vec{N}}q_{ij}) = \frac{2\kappa}{\sqrt q}G_{ijkl}{\tilde\pi}^{kl}$. Taking the trace yields
\begin{eqnarray}\label{GRq}
 \frac{1}{3}Tr{(K}) =\frac{1}{2N}\big(\frac{\partial\qt}{\partial t} -  \pounds_{\vec{N}}\qt\big)
= \frac{2\kappa\beta}{\sqrt q}{\bar H},
\end{eqnarray}
wherein the constraint  $(\beta \pi +{\bar H})=0$ has been used to arrive at the last step. Equation (\ref{GRq})
  demonstrates that the lapse function and intrinsic time are precisely related ({\it a posteriori} by the EOM) by the same formula as in (\ref{lap}). For a theory with full 4D diffeomorphism invariance (such as Einstein's GR with $\beta^2=1/6$ and consistent Dirac algebra of constraints), this relation is an identity that does not compromise the arbitrariness of $N$. However, it reveals (even in Einstein's GR) the physical meaning of the lapse function and its relation
to the intrinsic time.

\subsection{2.3 Paradigm shift and resolution of the problem of time}

Starting with only spatial diffeomorphism invariance and through constructive interference, Eq. (\ref{qeq}) with physical evolution in intrinsic time generated by $\bar H$ is obtained.
  This relates the momentum to the coordinate time derivative of the metric precisely as in (\ref{lap}).  It is thus possible to interpret the emergent classical spacetime (which can generically be described with an ADM metric) as possessing extrinsic curvature that corresponds precisely to the derived lapse function displayed in (\ref{lap}).
  However, {\it only} the freedom of spatial diffeomorphism invariance is realized, as the {\it emergent} lapse is now completely described by the intrinsic time $\qt$ and $\vec N$. All EOM w.r.t coordinate time $t$ generated by $\int NH +N^iH_i$ in Einstein's GR can be recovered from evolution w.r.t. $\qt$ and generated by $\bar H/\beta$  iff $N$ assumes the form of (\ref{lap}). All the previous observations lead to the central revelation: full 4D spacetime covariance (with its consequent baggage of arbitrary lapse and gauged histories) is a red herring that obfuscates the physical reality of time, and all that is necessary to consistently capture the classical physical content of GR is a theory invariant only w.r.t. spatial diffeomorphisms accompanied by a master constraint that enforces the dynamical content. The paradigm shift points to a {\it complete resolution of the problem of time}, from quantum to classical GR: classical spacetime, with consistent lapse function and ADM metric,
\begin{equation}\label{st}
ds^2=-\Big[\frac{(\partial_t \qt- \pounds_{\vec{N}}\qt)dt}{4\beta\kappa (\bar{H}/{\sqrt q})}\Big]^2+{q}_{ij}[dx^i+N^i dt][dx^j+N^j dt],
\end{equation}
emerges from constructive interference of a spatial diffeomorphism invariant quantum theory with Schr\"{o}dinger and HJ equations first order in intrinsic time development.
 Gratifying too are the correlations of classical proper time $d\tau$ and quantum intrinsic time $\qt$  (through
$d\tau^2 = [\frac{ \delta \qt}{(4\beta\kappa {\bar H}/{\sqrt q})}]^2$ (for vanishing shifts)), and of Wheeler's notion of ADM ``simultaneity" to quantum simultaneity ($dt=0 \Rightarrow \delta \ln q =0$  by Eq. (\ref{lap})).
In particular, by (\ref{eq:qdot}) and (\ref{GRq}), proper time intervals measured by physical clocks in spacetimes that are solutions of Einstein's equations always agree with the result of Eq. (\ref{st}).\\

In conventional formulations, 4-covariance is the underlying paradigm, and it has been argued that the chosen time variable should be a spacetime scalar. It follows that no function of the intrinsic geometry can fulfill this criterion. ``Scalar field time" has thus been advocated notably by Kuchar, and others, despite the fact that such a choice would not be available in the context of pure gravity. The novel and crucial feature is that the paradigm shift to only spatial diffeomorphism invariance is just what is needed to permit a degree of freedom constructed from the intrinsic spatial geometry to act as the time variable. Indeed, $\delta\ln q^{\frac{1}{3}} =\frac{1}{3}(\delta q)/q $, which is a scalar under spatial diffeomorphisms, is the apposite choice of intrinsic time interval.

\subsection{2.4 Improvements to the quantum theory}

The framework of the theory also prompts improvements to $\bar H$. The requirement of a real physical Hamiltonian density $\bar H$  compatible with spatial diffeomorphism symmetry suggests supplementing the kinetic term in the square root with a positive semidefinite quadratic form, i.e.
 \begin{eqnarray}\label{QH}
 \bar{H} =\sqrt{ {\bar G} _{ijkl}{\bar\pi}^{ij}{\bar \pi}^{kl} +
 [\frac{1}{2}(q_{ik}q_{jl} +q_{jk}q_{il}) +\gamma q_{ij} q_{kl}]\frac{\delta W}{\delta { q}_{ij}}\frac{\delta W}{\delta {q}_{kl}}} \nonumber \\ =\sqrt{[ {\bar q}_{ik}{\bar q}_{jl} +\gamma{\bar q}_{ij}{\bar q}_{kl}]({\bar\pi}^{ij}{\bar \pi}^{kl}
 + q^{\frac{2}{3}}\frac{\delta W}{\delta { q}_{ij}}\frac{\delta W}{\delta {q}_{kl}})}
 .
  \end{eqnarray}
%
%
${\bar H}$ is then real if $\gamma > -\frac{1}{3}$. To lowest order for perturbative power-counting renormalizabilty\cite{Horava}, $W = \int\,\big[ {\sqrt q}(a R - \Lambda) + g{\tilde\epsilon}^{ikj}(\Gamma^l_{im} \partial_j\Gamma^m_{kl} +\frac{2}{3}\Gamma^l_{im}\Gamma^m_{jn}\Gamma^n_{kl})\big]$ (i.e. of the form of a 3D  Einstein-Hilbert action with cosmological constant supplemented by a Chern-Simons action with dimensionless coupling constant $g$).
The potential of the form of Einstein's theory with cosmological constant is recovered at low curvatures.
 The effective value of $\kappa$ and cosmological constant can thus be determined as ${\kappa}=\frac{8\pi G}{c^3} = \sqrt{\frac{1}{2a\Lambda(1+3\gamma)}}$ and $\Lambda_{\rm eff}= { \frac{3}{2}\kappa^2}\Lambda^2(1+3\gamma)=\frac{3\Lambda}{4a}$ respectively. The possibility of having a new parameter $\gamma$ in the potential (different  from $\lambda$ in the supermetric) has been overlooked in previous works. Furthermore, positivity of $\bar{H}^2$ (with $\gamma > -\frac{1}{3}$) is correlated with {\it real} $\kappa$ and {\it positive} $\Lambda_{\rm eff}$.
 There is also the intriguing feature that the lowest classical energy of the physical Hamiltonian $\bar H$ occurs when zero modes are present i.e. $\gamma\rightarrow -\frac{1}{3}$, leading, in this limit and for fixed $\kappa$, to $\Lambda_{\rm eff}\rightarrow 0$. This, however, requires a thorough investigation of the renormalization group flow of $\gamma$ and other parameters to deduce the exact behavior of $\Lambda_{\rm eff}$ with physical energy scale, especially when matter and other forces are also taken into account.
A slight generalization  of (\ref{QH})  is to replace $\frac{\delta W}{\delta { q}_{ij}}$ in the positive semidefinite quadratic form with ${\sqrt q}(\Lambda' q^{ij}+ a' {R}q^{ij} + b{R}^{ij}+ g'C^{ij})$, which is the most general symmetric second-rank tensor (density) containing up to third derivatives of the spatial metric. The coupling constant $g'$ associated with the Cotton-York tensor $C^{ij}$ is dimensionless (${\sqrt q}C^{ij}$ is proportional to the functional derivative with respect to the spatial metric  $q_{ij}$ of the Chern-Simons term).

It must be pointed out that the Hamiltonian density above involves a square root and its rigorous definition through spectral decomposition remains a most formidable challenge. While this is not necessarily a defect of the theory,
issues in quantum gravity that involve the rigorous non-perturbative definition of the Hamiltonian operator are thus not yet addressed in the current work. There are some intriguing and encouraging signs. Out of the many possible alternatives that respect 3-covariance (rather than the more restrictive 4-covariance of Einstein's theory), Horava's power-counting renormalizable ``detailed balance" form of the Hamiltonian constraint \cite{Horava} (or its slight generalization discussed above which also yields a dimensionless coupling constant in the highest-order term $C^{ij})$ is essentially pinned down by the square root and positivity of the Hamiltonian density.

 In Ref.\cite{Horava}, the Hamiltonian density with ``detailed balance" is proportional to
$\frac{G_{ijkl}}{\sqrt q}({\pi}^{ij}{\pi}^{kl} +\frac{\delta W}{\delta { q}_{ij}}\frac{\delta W}{\delta {q}_{kl}})$. Without factoring out $\pi$,  the negative mode in the full supermetric, $G_{ijkl}$, compromises the positivity of the kinetic term, $\frac{G_{ijkl}}{\sqrt q}{\pi}^{ij}{\pi}^{kl}$, in the theory.  Formulations that use an extra scalar field\cite{Isham}, or variables other than $\qt$ as time to attain deparametrization, will also be afflicted with the same problem. In contradistinction, with $\ln q^{\frac{1}{3}}$ as intrinsic time, $\pi$ is singled out and isolated as the  conjugate variable, with the upshot that, in the Schr\"{o}dinger equation, ${\bar H}$ (which does not contain $\pi$) always has positive semidefinite kinetic term.

\subsection{2.5 Gauge-invariant global time, superspace dynamics, and temporal order}

 In our intrinsic time formulation, the Wheeler-DeWitt equation, which is a constraint that must be satisfied at each spatial point $x$, is replaced by a Schr\"{o}dinger equation with  $\frac{{\bar H}(x)}{\beta}$ generating translations in $\qt(x)$ which is a Tomonaga-Schwinger\cite{Tomonaga, Schwinger} many-fingered time variable. The transcription from, apparently many-fingered dynamics to evolution with respect to a gauge-invariant global time and the Heisenberg picture is, remarkably, unequivocal for 3-hypersurfaces which are compact Riemannian manifolds without boundary. Hodge decomposition of the 0-form $\delta\qt$ uniquely yields $\delta\qt   = \delta h + \nabla_i {\delta V}^i $, wherein $\delta h$  is harmonic, {\it independent} of $x$, and gauge-invariant, whereas $\delta V^i$ can be gauged away because, fortuitously,  $\pounds_{\delta N^i} \qt = \frac{2}{3}\nabla_i\delta N^i$.  This leads,  bearing in mind (\ref{HJ}), to the  transcription
 \begin{equation}
 i\hbar\frac{\delta \Psi}{\delta h} = \int i\hbar\frac{\delta \Psi}{\delta\qt(x) }\frac{\delta \qt(x)}{\delta h}d^3x=\left[\int\frac{{\bar H}(x)}{\beta}d^3x\right]\Psi \label{WE},
 \end{equation}
 which describes evolution with respect to the intrinsic superspace time interval $\delta h$. The corresponding physical Hamiltonian ${\mathcal H}_{\rm phys.}:=\int\frac{{\bar H}(x)}{\beta}d^3x$  is, moreover, spatial diffeomorphism invariant as it is the integral of a tensor density of weight one. This  remarkable Schr\"{o}dinger equation, $i\hbar\frac{\delta \Psi}{\delta h} ={\mathcal H}_{\rm phys.}\Psi$, dictates quantum geometrodynamics in {\it explicit}  superspace  $^{(3)}\mathcal{G}$ entities  ($ \Psi[[q_{ij}] \in {^{(3)}\mathcal{G}}],{\mathcal H}_{\rm phys.}, \delta h$).

 On equating $\delta t = {\mathcal K}\delta h$ in (\ref{st}), the emergent classical spacetime from constructive interference under superspace intrinsic time evolution will, as demonstrated in earlier subsections, be described by the ADM metric,
\begin{equation}
ds^2=-\Big[\frac{\delta h- \pounds_{{\vec{\mathcal N}\delta h}}\qt}{4\beta\kappa(\bar{H}/{\sqrt q})}\Big]^2+{q}_{ij}[dx^i+{\mathcal N}^i\delta h][dx^j+{\mathcal N}^j\delta h],
\end{equation}
 wherein ${\mathcal N}^i:= N^i/{\mathcal K}$.  Since it can always be absorbed into the gauge parameter $N^i$,  the constant ${\mathcal K}$ has no physical implication. Rather, it is the (scalar function) ${\kappa\bar H}/{\sqrt q}$ that provides the physical conversion between the dimensionless intrinsic time interval $\delta h$  and the proper time of $ds^2$. Gravitational redshifts and other physical effects are thus determined by the Hamiltonian density ${\bar H}$; and the proper time (with vanishing shifts and $dx^i=0$) $d\tau = \frac{\delta h {\sqrt q}}{4\beta\kappa \bar H}$ exhibits the physically intuitive property of varying directly with intrinsic superspace time interval and reciprocally with energy density.

  The crucial time development operator can be derived by integrating the Schr\"{o}dinger equation. This is now feasible without ambiguity because $\delta h$ is ``1-dimensional", more precisely, $x$-independent, {\it rather than many-fingered}. Moreover, the necessity of ``time"-ordering, which underpins the notion of causality, emerges because quantum fields do not commute at different ``times".  Equation (\ref{WE}) implies $\delta\Psi= [-\frac{i}{\hbar}{\mathcal H}_{\rm phys.}]\delta h\Psi$; thus yielding $\Psi[[q_{ij}(h)]\in {^{(3)}\mathcal{G}}] = U(h,h_0)\Psi[[q_{ij}(h_0)]\in {^{(3)}\mathcal{G}}]$, with $h$-ordered evolution operator
 $U(h,h_0):=T\exp\left[-\frac{i}{\hbar}\int^h_{h_0} {\mathcal H}_{\rm phys}(h')\delta h'\right]$. As ${\mathcal H}_{\rm phys.}$  is classically real and gauge invariant, a unitary and diffeomorphism-invariant $U(h,h_0)$ is viable. Since $\delta h$  is also unchanged under spatial diffeomorphisms, the temporal ordering  in $U(h, h_0)$ is reassuringly gauge invariant.

\section{3. Further discussions}

 That there are, for pure gravity, two physical degrees of freedom can be ascertained in the following way: spatial diffeomorphism invariance constrains the physical momenta ${\tilde\pi}^{ij}$ to be transverse ($\nabla_i {\tilde\pi}^{ij}=0$) leaving 3 remaining degrees. The 2 transverse traceless modes can be obtained through $\pi^{ij}_{TT}= ({\tilde\pi}^{ij}- \frac{q^{ij}}{3}\pi)-(\nabla^{i} W^{j}+\nabla^{j} W^{i} -\frac{2q^{ij}}{3}\nabla_kW^k) $, with $W^i$  the solution for $\nabla_i {\pi}^{ij}_{TT}=0$.
 Substituting this decomposition of ${\tilde\pi}^{ij}$ into the symplectic potential and integrating by parts terms with $W^i$ reveal that
  \begin{equation}\label{DOF}
 \int\, {\tilde\pi}^{ij}\delta q_{ij}
 = \int \left( {\pi}^{ij}_{TT}q^{\frac{1}{3}}\delta{\bar q}_{ij}- 2W^jq^{\frac{1}{3}}\nabla^i\delta{\bar q}_{ij} + \pi\delta\ln q^{\frac{1}{3}} \right),
\end{equation}
which yields $ \int \,({\bar\pi}^{ij}_{T}\delta{\bar q}^{{\rm phys.}}_{ij}+ \pi \delta \qt)$
when restricted to the physical subspace with $\nabla^i\delta{\bar q}^{\rm phys.}_{ij} =0$ (this condition has the geometrical meaning physical, $\delta{\bar q}^{\rm phys.}_{ij}$, and gauge, $\delta{\bar q}^{\rm gauge}_{ij}=  \pounds_{\vec{N} }{\bar q}_{ij}$, directions are orthogonal w.r.t. the spatial supermetric ${\bar G}^{ijkl}$).
This decomposition yields 2 physical canonical degrees of freedom $({\bar q}^{{\rm phys.}}_{ij},{\bar\pi}^{ij}_{T}:={\pi}^{ij}_{TT}q^{\frac{1}{3}})$, and
an extra pair $({\qt}, \pi)$ to play the role of time and Hamiltonian (which is consistently tied to $\pi$  by the dynamical equations (\ref{WD}) and (\ref{HJ})).
For perturbations about any background $q^*_{ij}= q_{ij}-\delta q_{ij}$, the linearized physical spatial metric modes $\delta{\bar  q}^{\rm phys.}_{ij} =
(P^{kl}_{ij})^*\delta q_{kl} $ are traceless ($q^{*ij}\delta{\bar q}^{\rm phys.}_{ij} =0$) and transverse
(${\nabla}^ {*i}\delta {\bar q}^{\rm phys.}_{ij} =0)$ w.r.t $q^*_{ij}$, correctly accounting for the perturbative graviton degrees of freedom.

With regard to Lorentz invariance, it should be pointed out that, although there is only spatial diffeomorphism invariance, for any classical ADM spacetime the Lorentz symmetry of the tangent space is intact, as the ADM metric $ds^2 = \eta_{AB}E^A\,_\mu E^B\,_\nu dx^\mu dx^\nu = -N^2dt^2 + q^{\frac{1}{3}}\bar{q}_{ij}(x,t)(dx^i + N^i dt)(dx^j + N^j dt)$ is invariant under local Lorentz transformations of the vierbein fields $e'^A_\mu =\Lambda^A\,_B(x)e^B\,_\mu$  which do not affect metric components $g_{\mu\nu} =\eta_{AB}e^A\,_\mu e^B\,_\nu$. This local gauge symmetry is associated with the freedom to choose  (up to Lorentz transformations) different tangent spaces at different spacetime points, rather than with the isometry of the metric under spacetime coordinate transformations.  A generic spacetime may have no isometry at all. At the more fundamental level of spatial metric,  it is possible to define\cite{Ashtekarbook} $q^{ij}:= \frac{1}{(\det{\tilde E})}{\tilde E}^{ai}{\tilde E}_a\,^j$, wherein the densitized triad ${\tilde E}^{ai}=\frac{1}{2}\epsilon^{ijk}\epsilon^{abc}e_{bj}e_{ck}= (\det e)[e_{ai}]^{-1}$ .
Both $q^{ij}$ and $q_{ij} := e_{ai}e^a\,_j$, are invariant under local (anti-)self-dual $SO(3,C)$ rotations of the triad, ${E}^{ai}$, and its inverse $e_{ai}$.  Moreover, expressing the spatial metric as $q_{ij}=e_{ai} e^a\,_j$  leads to the decomposition of the  symplectic potential as
 \begin{equation}\label{DOF}
 \int\, {\tilde\pi}^{ij}\delta q_{ij}
 = \int {\pi}^{ai}\delta e_{ai} =\int \left(\pi\delta\qt + {\bar\pi}^{ai}\delta{\bar e}_{ai}\right);
\end{equation}
wherein $\pi^{ai} := 2{\tilde\pi}^{ij}e^a\,_j, \pi = q_{ij}{\tilde\pi }^{ij} =\frac{1}{2}( \pi^{ai}e_{ai}), {\bar\pi}^{ai}:= e^{\frac{1}{3}}[\pi^{ai}- E^{ai}\frac{(\pi^{bk}e_{bk})}{3}]$, and unimodular
${\bar e}_{ai} := e^{-\frac{1}{3}}e_{ai}$ ,
together with the Gauss law constraint, $\frac{1}{2}[e^a\,_{i} \pi^{bi}-e^b\,_i\pi^{ai} ] \approx0$, which generates $SO(3, C)$ (isomorphic to $SO(3,1)$) Lorentz transformations of the variables. The conjugate pair $(\qt :=\ln e^{\frac{2}{3}},  \pi :=\frac{1}{2}( \pi^{ai}e_{ai}))$ is Lorentz invariant and commutes with the remaining variables $({\bar \pi}^{ai}, {\bar e}_{ai})$.

The inclusion of matter and other forces is rather straightforward as Standard Model fields do not couple to $\pi$ and the corresponding Hamiltonian density of these fields, $H_M$, can be appended to gravitational kinetic and potential energy terms. Consequently ${\bar H}_T= \sqrt{\bar H^2 + H_M}$  replaces ${\bar H}$ of (\ref{QH}), and the extension does not affect the fundamental form of the dynamical equation which is now $\beta\pi +\bar H_T =0$.
In $H_T$, Weyl fermions couple to $SL(2,C)$ spin connections.  For any classical ADM spacetime, these spin connections which couple to Standard Model fermions are pullbacks of the self and anti-self-dual Lorentz spin connections to 3-dimensional spatial slices\cite{Ashtekar, Soo}.

Identification of a complete set of observables in theories with diffeomorphism invariance is often thought to be more than a challenging task. In this context,
for a theory with the HJ equation first order in time, the solution is complete\cite{Courant,Landau} in that it has as many integration constants (denoted earlier by $\alpha$) as number of degrees of freedom in the theory, plus an overall additive constant. These are all gauge invariant and, together with $\omega :=\frac{\delta S}{\delta \alpha}$ (which expresses the coordinates in terms of time and the constants $(\alpha, \omega)$),  provide general integrals of equations of motion that are well suited to the role of physical observables of diffeomorphism-invariant theories. Alternatively, emergent spacetime manifolds obtained by integrating Hamilton's equations are characterized by $4\times\infty^3$ freely specifiable initial data $({\bar q}^{{\rm phys.}}_{ij},{\bar\pi}^{ij}_{TT})$. In a theory with only spatial diffeomorphism gauge symmetry, physical observables are required to commute only with $H_i$ (and not $H$), thus all Kuchar observables\cite{Kuchar} {\it become} physical. These observables will also consistently have weakly vanishing Poisson bracket with $\mathrm{M}$ (since
 $\{f(q_{ij}, {\tilde\pi}^{ij}), {\mathrm{M}} \}|_{\mathrm{M}=0
\Leftrightarrow H=0} \approx 0$).

The symplectic 1-form $ \int\, \pi \delta \qt = \int \frac{2}{3}(\frac{\pi}{\sqrt q})\delta{\sqrt q}$ allows a different perspective.
With the further restriction of $\nabla_i\pi=0 \Leftrightarrow \frac{\pi}{\sqrt q} =T$, York\cite{York1, York2}  interprets and deploys the scalar $\frac{\pi}{\sqrt q} =-\frac{Tr K}{6\beta^2\kappa} $  as the ``extrinsic time" variable.  It follows that the Hamiltonian $\bar H =-\beta\pi $ is then proportional to ${\sqrt q}$, and the total energy to the volume. In our framework, York's restriction of spatially constant $\frac{\pi}{\sqrt q} =-\frac{{\bar H}}{\beta\sqrt q}= T$ is a special case wherein, with vanishing shift vectors,
$d\tau^2=[{\frac{\delta h}{4\beta^2\kappa {T}}}]^2$.
Although the extrinsic time variable is then invariant under spatial diffeomorphisms,  it is, however, not invariant under 4D coordinate transformations, which are supposedly symmetries of Einstein's theory. With the paradigm shift to just spatial diffeomorphism invariance, $\delta\qt$ is well suited to the role of physical time interval: it is a spatial diffeomorphism scalar with a gauge-invariant part $\delta h$, which is spatially constant.

 From the Schr\"{o}dinger and HJ equations, pure general relativity has the physical content of conjugate variables $({\bar q}_{ij},{\bar \pi}^{ij})$ subject to $H_i=0$
 evolving w.r.t. $\ln q^{\frac{1}{3}}$ with effective Hamiltonian density ${\bar H}/\beta$. Thus proceeding from the action
$\int\,[{\bar \pi}^{ij}{\delta{\bar{q}}_{ij}} - \frac{\bar{H}}{\beta}\delta\ln q^{\frac{1}{3}}]- \int \delta{\mathcal N}^iH_i$, and inverting for ${\bar \pi}^{ij}$ in terms of $\frac{\delta{\bar{q}}_{ij}}{\delta\ln q}$ from the EOM, yields (the result can also be deduced from (\ref{qeq}) and (\ref{p})) the action functional as
\begin{equation}
S=-\int\sqrt{V}\sqrt{\frac{1}{\beta^2}(\delta\ln q^{\frac{1}{3}} -\pounds_{\delta{\vec{\mathcal N}}}\ln q^{\frac{1}{3}})^2
-{\bar G}^{ijkl}(\delta {\bar q}_{ij} -\pounds_{\delta{\vec {\mathcal N}}}{\bar q}_{ij})(\delta {\bar q}_{kl} -\pounds_{\delta{\vec{\mathcal N}}}{\bar q}_{kl})},
\end{equation}
 which is just the superspace proper time with $\sqrt V =\sqrt{{\bar H}^2 -{\bar G}_{ijkl}{\bar\pi}^{ij}{\bar\pi}^{kl}}$ playing the role of ``mass" if it were constant.
 This regains the generalized Baierlein-Sharp-Wheeler action\cite{BSW} which has also been studied in Ref.\cite{Barbour, Anderson2} in a different situation.

Transparent and consistent dynamics revealing the primacy of the physical Hamiltonian ${\mathcal H}_{\rm phys.}$ and the role of intrinsic time in general relativity and its extensions (a related discussion on the initial data formulation can be found in Ref. \cite{Niall}) can be obtained from several complementary approaches: the master constraint formulation which recovers the correct physical content from the usual starting point of canonical general relativity; the Schr\"{o}dinger equation (4) (or its superspace version (14)) as the fundamental equation for quantum geometrodynamics; the generalized  Baierlein-Sharp-Wheeler action (18); and also, perhaps most important to a causal quantum theory, the evolution operator $U(h, h_0)$ with gauge-invariant temporal ordering.

The extension from classical to quantum theory is dependent on operator ordering and radiative corrections. Although ${\bar G} _{ijkl}{\bar\pi}^{ij}{\bar \pi}^{kl}$ in (\ref{QH}) is naturally associated with (the negative of) the Laplacian operator $\frac{\delta}{\delta {\bar q}^{ij}}\bar G_{ijkl}\frac{\delta}{\delta {\bar q}^{kl}}$ , it must be pointed out that issues in quantum gravity that depend on the rigorous definition and regularization of the Hamiltonian operator are not yet addressed in this work.

\section*{Acknowledgements}

This work was supported in part by the National Science Council of Taiwan under Grant No.
NSC101-2112-M-006 -007-MY3; the Institute of Physics, Academia Sinica; and the National Center for Theoretical
Sciences, Taiwan.


\appendix
\section{Appendix}

The method of master constraint can be used to eliminate fictitious symmetry without losing the physical content. The example of the simple relativistic point particle is quite instructive. The action is
\begin{eqnarray}\label{lag}
S&=& -m_0c \int \sqrt{-\eta_{\mu\nu}dx^\mu dx^\nu} = -m_0c^2\int d\tau\nonumber\\
 &=& -m_0c^2\int dt \sqrt{1-(\frac{d\vec{x}}{cdt})^2} =\int (\vec{p}\cdot\frac{d\vec{x}}{dt}- c \sqrt{\vec{p}^2 +m_0^2c^2}) dt,
\end{eqnarray}
with $\vec{p}=m_0\frac{d\vec{x}}{d\tau}$, and the physical Hamiltonian $\bar H=c \sqrt{\vec{p}^2 +m_0^2c^2}$  emerges when $t$ is correctly identified as the time variable. On the other hand, in the `manifestly covariant' approach, introduction of an extraneous `time' parameter $\lambda$ results in
\begin{equation}
S = -m_0c\int \sqrt{-\eta_{\mu\nu} \frac{dx^\mu}{d\lambda} \frac{dx^\nu}{d\lambda} }d\lambda = \int {\mathcal L}[x^\mu,\frac{dx^\mu}{d\lambda} ]d\lambda \quad\Rightarrow p_\mu =\frac{\partial {\mathcal L}}{\partial (\frac{dx^\mu}{d\lambda})} = m_0\eta_{\mu\nu}\frac{dx^\nu}{d\tau},\nonumber
\end{equation}
with $p_\mu  \frac{dx^\mu}{d\lambda} - {\mathcal L} =0$ i.e. exactly vanishing Hamiltonian; but the momenta are constrained by  $H=p^\mu p_\mu +{m_0}^2c^2 = -(p_0 -\frac{\bar H}{c})(p_0 +\frac{\bar H}{c})=0$. Formulating the constrained theory with
\begin{equation}
S =\int  [p_\mu \frac{dx^\mu}{d\lambda}- N(p^\mu p_\mu + {m_0}^2c^2)]d\lambda
\end{equation}
 results in EOM which (a posteriori) determine $Nd\lambda = \frac{cdt}{2p^0}=\frac{cdt}{(2\sqrt{\vec{p}^2+{m_0}^2c^2})} = \frac {c^2dt}{2{\bar H}}$. The obfuscating reparametrization `symmetry' associated with $\lambda$ (which is not intrinsic to the theory) and constrained Hamiltonian $N(p^\mu p_\mu + {m_0}^2c^2)$ gives rise to artificial gauge histories of $x^\mu$ in $\lambda$-time with Lagrange multiplier $N$.
Correctly isolating one of the degree $x^0$ as the intrinsic time, and forgoing the `$\lambda$-reparametrization symmetry' regains the much more transparent  description of (\ref{lag}) with dynamical variables $\vec x$  evolving w.r.t. intrinsic time $t = \frac{x^0}{c}$ and physical Hamiltonian $\bar H$.  In the event (A2) is the starting point (analogous to the situation in General Relativity), the physical description of (A1) can be recovered by introducing the master constraint $M=(p_0 +\frac{\bar H}{c})^2 =0$ (which makes $H=0$ redundant and in effect replaces it); thus yielding
\begin{eqnarray}
S&=&\int [p_\mu \frac{d x^\mu}{d\lambda}] d\lambda - \int \textsl{m}(p_0 +\frac{\bar H}{c})^2  d \lambda \nonumber \\
&=&\int [p_0 c +\vec{p} \cdot \frac{d\vec{x}}{dt}]dt -\int {\textsl{m}}(p_0 + \sqrt{\vec{p}^2+ {m_0}^2c^2})^2 d \lambda \nonumber\\
&=&\int [-c\sqrt{\vec{p}^2+ {m_0}^2c^2} + \vec{p} \cdot {\frac{d\vec{x}}{dt}}]dt  +\int \textsl{m}(p_0 + \sqrt{\vec{p}^2+ {m_0}^2c^2})^2 d \lambda,
\end{eqnarray}
which implies $\bar H=c\sqrt{\vec{p}^2+ {m_0}^2c^2}$ is the effective Hamiltonian for the variables $(\vec{x},\vec{p})$, and the $M$ constraint is equivalent to $p_0 =-\frac{\bar H}{c}$ which can consistently be interpreted classically as the Hamilton-Jacobi equation $\frac{\partial S}{\partial x^o} + \frac{\bar H}{c} =0$, and quantum mechanically as a Schr\"{o}dinger equation.

\end{document}